\documentclass[preprint,aps]{revtex4}

\usepackage{graphicx}

\begin{document}
%
\title{Development of a Vacuum Ultra-Violet Laser$-$Based Angle$-$Resolved Photoemission System with a Super-High Energy Resolution Better Than 1 meV}
%
%
\author{Guodong Liu$^{1}$, Guiling Wang$^{2}$, Yong Zhu$^{3}$, Hongbo Zhang$^{2}$,
Guochun Zhang$^{3}$, Xiaoyang Wang$^{3}$, Yong Zhou$^{2}$, Wentao
Zhang${^1}$, Haiyun Liu$^{1}$,Lin Zhao$^{1}$, Jianqiao Meng$^{1}$,
Xiaoli Dong$^{1}$, Chuangtian Chen$^{3}$, Zuyan XU$^{2}$ and X. J.
Zhou $^{1,*}$} \affiliation{
\\$^{1}$National Laboratory for Superconductivity, Beijing National Laboratory for Condensed Matter Physics, Institute of Physics, Chinese Academy of Sciences, Beijing 100080, China
\\$^{2}$Laboratory for Optics, Beijing National Laboratory for Condensed Matter Physics,Institute of Physics, Chinese Academy of Sciences, , Beijing 100080,  China
\\$^{3}$Technical Institute of Physics and Chemistry, Chinese Academy of Sciences, Beijing 100080, China}

\date{Oct.16, 2007}

\begin{abstract}

 The design and performance of the first vacuum ultra-violet (VUV) laser-based angle-resolved photoemission (ARPES) system are described. The VUV laser with a photon energy of 6.994 eV and bandwidth of 0.26 meV is achieved from the second harmonic generation using a novel non-linear optical crystal KBe$_2$BO$_3$F$_2$ (KBBF).  The new VUV laser-based ARPES system exhibits superior performance, including super-high energy resolution better than 1 meV, high momentum resolution, super-high photon flux and much enhanced bulk sensitivity, which are demonstrated  from measurements on a typical Bi$_2$Sr$_2$CaCu$_2$O$_8$ high temperature superconductor. Issues and further development related to the VUV laser-based photoemission technique are discussed.

\end{abstract}


\maketitle



\section{Introduction}


Angle-resolved photoemission spectroscopy (ARPES) is a powerful tool to probe the electronic structure of solids\cite{Huefner,SKevan}. When a light illuminates on a material, it kicks out photoelectrons, a phenomenon well known as the photoelectric effect\cite{EinsteinPE}. ARPES measures the energy distribution of photoelectrons at different emission angle with respect to the sample surface normal, thus directly measuring two fundamental quantities for describing the electronic state of a solid, i.e., energy (E) and momentum (k). Theoretically, under the sudden approximation, ARPES measures a single particle spectral function A(k,$\omega$) with k and $\omega$ representing momentum and energy respectively\cite{SFunction}, a quantity that can be proposed or calculated from theoretical models.  The power of ARPES in directly extracting electronic state information and its direct comparison between experiment and theory have made it a prime choice in studying advanced materials, particularly high temperature superconductors\cite{DamascelliReview,CampuzanoReview,ZhouReview} and other strongly correlated electron systems\cite{SpecialJESRP}.


Because most of the macroscopic properties of a material are dictated by its microscopic electron dynamics, specifically within an energy range of a few k$_{B}$T (k$_B$ being the Boltzman constant and T temperature) near the Fermi level, high energy resolution has been a long-sought goal of photoemission technique in order to probe intrinsic electronic structures\cite{Huefner}.  Started in the early 1970s\cite{NSmith}, ARPES has experienced a dramatic improvement in its instrumental performance, particularly in the last two decades that can be accounted for by a couple of reasons.  First, because it is an indispensable technique in band mapping of materials, there has been a strong impetus from  the stringent requirement of prominent scientific issues to push the ARPES technique forward. Among the numerous phenomena manifested by the correlated electronic materials, one of the most fascinating examples is high temperature superconductivity,  which has been at the forefront of condensed matter physics research since its discovery in 1986\cite{BednorzMuller}. Later on, collossal magnetoresistive materials and other exotic and extreme quantum phenomena in condensed matter physics have been discovered\cite{ScienceSpecial}. Second, the development benefits greatly from the synchrotron radiation technology, its development from the first generation to the contemporary third generation, has provided the photoemission technique a high performance light source with high energy resolution, high photon flux and photon energy tunability. Third, the advancement of photoemission technique has benefitted profoundly from the dramatic advancement of electron detection technology. With the advent of an advanced electron energy analyzer from Scienta in the middle 1990s, the energy resolution experienced a jump from long-staggering 20$\sim$40meV to 5 meV, which can now reach 1 meV or better for the latest electron analyzers\cite{ScientaWebSpecsWeb}. Particularly, the evolution of electron detection  from one-dimensional to two dimensional scheme has not only improved the angular resolution from previous 2 degrees to nearly 0.2 degrees, but greatly enhanced the efficiency in data acquisition and the way of ARPES data analysis\cite{MDCMethod}. It is now routinely possible to get an energy resolution around $\sim$10 meV and angle resolution of 0.2 degrees combining the latest synchrotron radiation light source and an advanced electron energy analyzer.



The latest development has elevated ARPES technique from a traditional band mapping tool to a probe of many-body effect, i.e., the interaction of electrons with other entities like electrons, phonons and etc. in a solid\cite{ZhouReview}.  Some prominent scientific issues, like mechanism of high temperature superconductivity, and potential necessity in developing  a new theoretical paradigm for strongly correlated electron systems, have put even higher standard on the photoemission technique. One is apparently the further improvement of energy and momentum resolution. Various energy scales on the order of 1meV, like a superconducting energy gap in conventional superconductors\cite{HuefnerPbShinPb}, some high temperature superconductors, and other recently discovered novel superconductors, ask for an energy resolution of ARPES system to be near 1 meV or better. The probing of many-body effects, specifically the mode coupling in the electron self-energy, also asks for such a super-high energy resolution\cite{ZhouPRL}.


One obvious characteristic of present ARPES technique is its extreme surface sensitivity. This is related to the short inelastic mean free path in the photon energy range of interest\cite{Seah}. Between the energy range of 20 and 50eV that is most commonly used for valence band photoemmsion, the corresponding inelastic mean free path is only on the order of 5$\sim$10 $\AA$, i.e., representing only the topmost one or two layers of the sample. While this surface sensitivity is beneficial for surface study, it may stand out as a serious drawback in probing bulk properties. For understanding problems of the bulk nature, like in high temperature superconductors, it has been a long-standing concern how much the photoemission measured results are representative of the bulk properties. To overcome this shortcoming, one approach is to move to the high photon energy which will enhance bulk sensitivity, e.g., at 1000 eV, the corresponding inelastic mean free path can be increased up to $\sim$20 $\AA$\cite{Suga}. However, this slight enhancement of bulk sensitivity is achieved at a sacrifice of signal intensity, lower energy resolution and poor momentum resolution\cite{Suga}.


Another way to enhance bulk sensitivity is to move to lower photon energy. Lower photon energy also makes it easier to achieve high energy resolution and gives rise to better momentum resolution even for the same angular resolution.  With these potential advantages, it seems to be a natural choice to develop laser-based photoemission technique and this idea indeed has been  around for a long time. However, a couple of requirements have to be satisfied before a laser can be used for high resolution photoemission: (1). High photon energy;  For photoemission process to occur, the photon energy must be large enough to overcome the work function of the material under measurement. Since a work function of 4$\sim$5 eV is common, a photon energy larger than 5 eV is essential. (2). High photon flux;  The laser to be used must have enough photon flux to get decent photoemission signal. (3). Narrow bandwidth;  The bandwidth of the laser will be a major factor in determining the overall instrumental energy resolution. To get high energy resolution, narrow bandwidth of laser is necessary.  (4). Continuous-wave (CW) or quasi-continuous wave laser; Because photoemission process involves the space charge effect that may shift the energy position and broaden the linewidth\cite{ZhouSCE}, it is preferable to use CW or quasi-CW laser in order to minimize the space charge effect for getting high energy resolution. (5). Compatible electron detector; The relatively low energy of photoelectrons puts stringent demand on the electron energy analyzer that can not only work at such a low electron kinetic energy, but also has an energy resolution that is good enough to take full advantage of the narrow bandwidth of the laser. Theoretically, there has been also a concern whether the sudden approximation that has been the theoretical foundation for treating high-photon energy photoemission process, is still valid at low photon energy\cite{SuddenAP}.

The recent progress in generating ultra-violet (UV) and vacuum ultra-violet (VUV) laser by utilizing non-linear optical crystals has made a big progress in laser photoemission\cite{ShinP,DessauRSI,NoteVUV}. Using KBe$_2$BO$_3$F$_2$ (KBBF) crystal, Shin's group developed a quasi-CW VUV laser with a photon energy (h$\nu$=6.994 eV) and applied to photoemission technique\cite{ShinP}. The system realized for the first time an energy resolution better than 1 meV, achieving an overall energy resolution of 0.36 meV, 0.26 meV from the laser source and 0.25 meV from the electron energy analyzer\cite{ShinP}. However, the photoemission measurements reported so far have been confined to angle-integrated mode\cite{ShinP}. Independently,  Dessau's group reported successful development of UV-laser based angle-resolved photoemission setup\cite{DessauRSI}. The highest photon energy they can achieve is 6.05 eV, with a linewidth of 4.7 meV, while the overall energy resolution of the system is $\sim$8meV. Compared with synchrotron light source, much improved performance has been observed through using this UV laser-based ARPES syetsm\cite{DessauRSI}. However, in addition to a relatively poor energy resolution compared to the VUV laser used by Shin's group\cite{ShinP}, the relatively low photon energy limits the momentum space that can be reached.  Particularly the inability to reach the antinodal region in high temperature superconductors and other related correlated electron systems poses as a serious limitation in this UV-laser based ARPES setup.


In this article, we report a VUV laser-based ARPES system that has not only a super-high energy resolution better than 1 meV, but also with angle-resolved mode that can cover larger momentum space; specifically it can approach the antinodal region in high temperature superconductors and other related materials. To our best knowledge, this is the first VUV-laser based ARPES system with an energy resolution better than 1 meV.  In the following sections, we will describe the design of the experimental VUV laser ARPES setup. We will also demonstrate its superior performance by some testing and measuring on a typical Bi$_2$Sr$_2$CaCu$_2$O$_8$ (Bi2212) high temperature superconductor.

\section{Experimental System Design}

Fig. 1 shows a photograph of our VUV laser-based ARPES system. As labeled in the figure, the setup is composed of two major parts: the VUV laser optical system and the angle-resolved photoemission spectrometer system. The latter is further divided into four main components: (1). mu-metal ARPES measurement chamber; (2). Scienta R4000 electron energy analyzer; (3). Sample manipulator and cryostat; and (4). Preparation chamber and sample transfer system. In the following, we will give more detailed description of these components individually.\\

\subsection{The VUV laser optical system}

A simple and straightforward way to generate UV and VUV laser is to utilize the higher harmonic effect of the non-linear optical (NLO) crystal. A good VUV NLO crystal should have a large NLO coefficient, short absorption edge below 200 nm and moderate birefringence (for practical use, the birefringence should be moderate, 0.07$\sim$0.10). Fig. 2 summarizes the shortest Second-Harmonic Generation (SHG) wavelength available for the typical and well-developed NLO crystals. Even though $\beta$-BaB$_2$O$_4$ (BBO) and LiB$_3$O$_5$ (LBO) are very useful for second harmonic generation from visible to ultraviolet all the way to 200 nm, their capabilities below 200 nm are severely limited. In the case of BBO, since the absorption edge is at 189 nm, it can only achieve SHG output not shorter than 210 nm\cite{BBORef}.  For LBO, although the transparent range is wide, the phase-matching is limited by its small birefringence. As a result, the shortest SHG output wavelength is only 276 nm under phase-matching conditions\cite{LBORef}.

KBe$_2$BO$_3$F$_2$ (KBBF) is a novel UV NLO crystal that can break the 200nm wavelength barrier and enter the VUV region\cite{ChenRef,Chen2002,Togashi,ParameterRef}.  It has a large NLO coefficient of d$_{11}$=0.49pm/V, good birefringence of  n=0.072 and a very short cut-off wavelength of 152 nm\cite{ParameterRef}. So far it is the only  crystal available that enables phase matching below 160 nm for SHG.  However, a major obstacle of using KBBF lies in its plate$-$like nature. It is difficult to grow large-sized KBBF crystal thicker than one millimeter. Such a thin crystal cannot be cut at the phase-matching angle. To solve this problem, Chen et al. developed a prism-coupling technique (prism-coupled technique device, PCT) as schematically shown in Fig. 3\cite{Chen2002,Togashi,ParameterRef}, which is composed of a KBBF crystal sandwiched in between two CaF$_2$ prisms with a proper apex angle. The two interfaces between KBBF and CaF$_2$ prisms were brought into optical contact, and they are pressed from both sides by two stainless-steel pieces with four screws to maintain long-term optical contact. At present, the transmission loss per interface is approximately 5$\%$.  By using this KBBF$-$PCT device,  172.5 and 163.3 nm wavelengths have been demonstrated which are the shortest achieved  so far by nonlinear crystals for second-harmonic generation (SHG) and sum-frequency mixing (SFM), respectively\cite{Togashi}.

The 177.3 nm VUV laser used in our ARPES system is generated by frequency doubling of the original 355 nm laser through the KBBF$-$PCT device; the related optical system is schematically shown in Fig. 4. The key element is an optically contacted KBBF-CaF$_2$ prism-coupled device (the thickness of KBBF is 1.2 mm). The KBBF-PCT device sits on a stage with XYZ translation and one rotation in order to adjust the position and phase matching angle. The KBBF-PCT stage, together with the two following reflection mirrors, are motor-controlled from outside of the optical chamber. The 355 nm laser comes from a  frequency$-$tripled Nd:YVO4 laser (Vanguard, Spectra Physics) which produced a train of 10-ps pulses with a repetition rate of 80 MHz and a maximum output power of 4 watts. It is introduced into the frequency doubling chamber by a pair of reflection mirrors and a focus lens. After frequency doubling through the KBBF-PCT device, the 177.3 nm VUV laser is reflected by two mirrors and focused onto the sample position in the measurement chamber.   Because the 177.3nm can easily be absorbed by air, the KBBF-PCF device and all the optical elements hereafter are put inside an optical chamber. The chamber is filled with inert gas like helium or nitrogen. A complete inert gas exchange is necessary or the remnant air in the optical chamber can significantly reduces the 177.3nm laser power. The optical chamber and the ARPES measurement chamber are separated by a CaF$_2$ window which can allow the 177.3nm laser pass through while keeping the ultrahigh vacuum of the measurement chamber intact. The 355nm and 177.3nm lasers have linear polarization with the E-vector pointing to the vertical direction.

The advantages of our 6.994eV VUV laser over 6 eV UV laser lie in both the superior energy resolution and larger momentum space coverage. The linewidth of our VUV laser is $\sim$0.26meV, while it is $\sim$5 meV for the UV laser\cite{DessauRSI}.  At h$\nu$=6.994 eV, the maximum in-plane momentum covered is 0.84 $\AA$$^{-1}$ taking a work function of 4.3 eV that is common for high-T$_c$ materials, while it is 0.67 $\AA$$^{-1}$ for h$\nu$=6 eV.  The advantage of larger photon energy is obvious here because VUV laser can reach the important ($\pi$,0) antinodal region of high-T$_c$ materials and other transition metal oxides (the corresponding momentum is $\pi$/a with a being the lattice constant. For high-T$_c$ materials, a$\sim$3.8 $\AA$, so ($\pi$,0) corresponds to a momentum of $\sim$0.83 $\AA$$^{-1}$).

\subsection{Advanced angle-resolved photoemission spectrometer}

The spectrometer system has been designed to be compact, efficient and user-friendly. It is composed of mu-metal measurement chamber, electron energy analyzer, sample manipulator and cryostat, and preparation chamber and sample transfer system.\\
\\
{\bf i. Ultra-high vacuum mu-metal measurement chamber}\\

In an angle-resolved photoemission measurement, the photoelectron emission angle carries important information about its momentum. To keep the trajectory of the photoelectrons intact, the residual magnetic field near the sample region and on the way to the electron analyzer must be minimized.  This is particularly crucial for laser-based photoemission system because photoelectrons with low kinetic energy are more susceptible to small magnetic field. To best screen the earth field or the magnetic field from surrounding environment, we have made our measurement chamber directly using mu-metal. In addition,  another mu-metal magnetic shield liner is added inside the mu-metal chamber and all the ports are connected with long mu-metal sleeves.  The liner is tightly coupled to the outside one of the two mu-metal layers of the electron energy analyzer. The measurement chamber and electron energy analyzer are demagnetized together after assembly. With all these delicate configurations, the remanent field in the sample region reaches a very low level with three components of 0.1, 0.5 and 0.6 mGauss, an overall magnitude of 0.8 mGauss.

The measurement chamber is connected to a cryopump and an ion pump for reaching ultra-high vacuum. After baking, it can reach a base pressure better than 5$\times$10$^{-11}$ Torr.  It is also equipped with a GAMMADATA high-flux discharging lamp with a toroidal grating monochromator which can provide UV light at 21.2 eV (helium I) and 40.8 eV (Helium II).\\
\\

{\bf ii. State-of-the-art electron energy analyzer}\\

The photoelectron signal is measured using a high-resolution hemispherical electron analyzer, GAMMADATA-SCIENTA R-4000.  One of the outstanding features of this analyzer is its high energy resolution that reaches the sub-meV region. For example, with a slit size of 0.1mm, and a pass energy of 1 eV, the energy resolution of the analyzer can approach 0.25 meV. This is well matched to the extremely narrow linewidth ($\sim$0.26 meV) of our VUV laser source to give an overall energy resolution better than 1 meV.

Another unique feature of the R4000 analyzer lies in its two-dimensional detection scheme that can  measure the intensity of photoelectrons over a wide angle range simultaneously. Since the kinetic energy of photoelectrons excited by our VUV laser is rather low, it was a big challenge to make the angular modes work at such low kinetic energies . The performance of the angular mode also depends sensitively on the spot size on the sample. Much effort has been put into optimizing the angular mode lens tables that satisfies the stringent requirements of low energy laser ARPES.  We have three angular modes that can cover 30 degrees, 14 degrees and 7 degrees and three sets of such lens tables for the spot size of 2.0mm, 0.8mm, and 0.1mm, respectively.  The 30 degree angular mode is particularly useful for covering large momentum space at the same time. \\
\\
{\bf iii.Sample manipulator and cryostat}\\

The sample manipulator consists of a three-axis positioning device for XYZ translation (VG: Centiax Translator) and a differentially pumped rotating platform (VG: RP100M) for the polar rotation (named hereafter $\varphi$ )of the sample.  The custom-made low temperature cryostat has two axis rotations of the sample: the tilt ($\theta$)and azimuthal ($\omega$) which is put on top of the rotary seal.  In this way, the sample has access to all 6-degrees of freedom including three translations and 3 rotations.
All the translations and rotations are driven by stepping motors (McMillan motion control Inc.) that are controlled through a computer.  A Labview program is developed to control the six-axis motions in a versatile way such as absolute movement, relative movement, origin adjustment, and position calibration and limitation. The control accuracy and reproducibility of the angles are 0.005 degree. The accuracy of X, Y and Z movements is 0.001, 0.001 and 0.01mm, respectively.

The cryostat is cooled down by continuous flow of  liquid helium and the lowest temperature the goniometer can reach is $\sim$12 K at the sample position. The cryostat has a button-heater embedded near the end of the cold-head that can vary the sample temperature between 12 and 450 K. Controlled by an advanced proportional-integral-derivative (PID) temperature controller, the sample temperature can be stabilized to be within 0.1 K. The temperature is measured by a standard Si-diode (LakeShore: DT-471) with a measurement accuracy of 1 K.  In order to minimize the sample position shift during sample temperature change due to thermal expansion/contraction, the goniometer is not attached directly onto the end of cryostat cold-end. Instead, it is mounted near the end of a stainless steel tube surrounding the cryostat rod. This supporting tube remains at room temperature so it does not experience expansion or contraction when the cryostat undergoes a warming/cooling process. The goniometer is connected to the cold-end of the cryostat through a flexible copper braid, a copper radiation shield is installed to reduce radiation heating of the sample stage. In order to thoroughly eliminate the stray magnetic field in the sample region, special caution is taken to assure that only non-magnetic components and materials are used in the goniometer and cryostat fabrication.\\
\\
{\bf iv. Prep chamber and sample transfer system}\\

Above the mu-metal measurement chamber is a sample preparation chamber which is separated from the measurement chamber by a UHV gate valve. This configuration makes it easy and time-saving for cryostat maintenance. The preparation chamber is equipped with a wobble stick for cleaving samples, a sputtering gun for cleaning sample, and a low-energy electron diffraction (LEED) for sample surface characterization. It also hosts a Residual Gas Analyzer (RGA) for leak checking. The preparation chamber is connected to the sample transfer system that can transfer samples in and out of the preparation chamber from air without baking or breaking ultrahigh vacuum in the preparation chamber. The sample transfer system consists of two stages separated by a valve. When transferring samples, the first stage is vented and loaded with samples. After pumping to a vacuum of 10$^{-7}$$\sim$10$^{-8}$ Torr, the samples can be transferred into the second stage with a better base pressure of 10$^{-9}$$\sim$10$^{-10}$ Torr, and further transferred to the preparation chamber.

\section{System Performance}

\subsection{Energy resolution test}

The energy resolution of the VUV laser-ARPES system is calibrated by using the Fermi edge of a cleaned polycrystalline gold, as shown in Fig. 5 measured using 2 eV pass energy and 0.1 mm analyzer slit of the electron energy analyzer. Here the gold is attached directly to the end of the cryostat end so it can reach a lower temperature of 9.22 K. By fitting the Fermi edge with a Fermi distribution function, the 12$\%$$\sim$88$\%$ width is obtained to be 3.211 meV. After subtracting the temperature broadening contribution (9.22K corresponding to 3.162 meV), one gets an overall energy resolution of the VUV-Laser ARPES system as 0.56 meV. This overall width is usually composed of three main contributions: (1). VUV laser linewidth (0.26meV); (2). analyzer resolution (for 2 eV pass energy and 0.1mm slit, an ideal energy resolution is 0.5 meV); (3). contribution of space charge effect\cite{ZhouSCE} which depends on photon flux, spot size, and other factors. In this case, the contribution from the laser source and the analyzer gives an energy resolution of 0.56 meV that comes very consistent with the measured value (0.56 meV). Note that, because of relatively high sample temperature, there is an uncertainty in getting precise energy resolution when trying to extract a small contribution from the main contribution of temperature. We did not use the best measurement condition, i.e., 1 eV pass energy and 0.1 mm slit of the analyzer, to test the energy resolution because of the uncertainty. In this case, the combined energy resolution from both the VUV laser (0.26 meV) and the analyzer (0.25 meV) would amount to 0.36 meV, as demonstrated by Shin's group tested using low sample temperature (2.9K)\cite{ShinP}.  We note that, although we can not test such high energy resolution because of our relatively high temperature of gold, the performance of our ARPES system is similar to that from Shin's group so the best energy resolution of 0.36 meV can be expected also in our system.

\subsection{Photon flux test}

The 177.3nm laser power generated right behind KBBF$-$PCT device is measured with a power meter (LP-3A, Physcience Opto-electronics Co., Ltd., Beijing), calibrated using a green laser at 532 nm (calibration at 177.3 nm is not available) by National Institute of Metrology of China. Using a KBBF-PCT with a KBBF thickness of 0.8 mm, a frequency-doubling power of 1.68 mW is achieved with a 2 Watts input power at 355 nm.  As the input power increases, or the thickness of KBBF increases, a higher 177.3 nm laser power is achievable.  Assuming the calibration of the power meter at 532 nm is close to that at 177.3nm, 1.68 mW corresponds to a 177.3 nm photon flux of 1.5$\times$10$^{15}$  photons$\slash$S. This photon flux is nearly two or three order of magnitude higher than that available from the third-generation synchrotron undulator beamlines.  In the mean time, the bandwidth of our VUV laser (0.26 meV) is more than one order of magnitude better than the synchrotron radiation (usual working resolution: 10$\sim$15meV, the best working resolution is $\sim$4 meV\cite{Hisor}). The 177.3 nm output power is stable within 5$\%$ drift in 8-hour period with less than $\pm$ 2$^{o}$C temperature change in the environment. Such a high stability is ideal for running an ARPES experiment. Due to heat accumulation and radiation damage, a local region ($\sim$ 100 micrometers) in the CaF$_2$ or KBBF crystals may get damaged, and the incident position of the 355nm laser on the KBBF$-$PCT device needs to be changed periodically.

\subsection{Angular mode test}

The performance of the angular mode is tested using a special Scienta wire-and-slit device which is composed of a thin wire and a set of slit. An electron gun hits the wire and the scattered electrons pass through the slit and are detected by the analyzer. The wire and slit are arranged in the way that the spacing between the adjacent two lines represents a known emission angle (in our case, 2.5 degrees).   Fig. 6  shows examples of the test patterns that are collected at 1eV, 2eV and 5 eV pass energies, respectively,  and 30 degree angular mode with 0.1mm spot size. It gives a constant dispersion over a wide kinetic energy range, with a kinetic energy as low as 0.5 eV. This makes it suitable for performing ARPES measurements using 6.994 eV VUV laser where the highest kinetic energy of the photoelectrons is nearly 2.7 eV with a common work function of 4.3 eV.

The angular resolution of the analyzer is also tested using the same wire-and-slit device. For 0.8 mm spot size, it is 0.3 degree for the 14 degree angular mode, while it is 0.8 degree for the 30 degree angular mode. Note that the angular resolution is sensitive to the spot size: the smaller the spot size, the better the angular resolution. Since it is not possible to do the angular resolution test for the 0.1 mm spot size using the same wire-and-slit scheme, the exact angular resolution is not known for our 0.1mm laser spot although the upper limit is estimated definitely much less than 0.8 degree for the 30 degree angular mode. Note that, since the momentum resolution $\Delta$k near the Fermi level is related to the angular resolution $\Delta$$\theta$ by  $\Delta$k=0.5118$\sqrt{h\nu-W}$cos$\theta$$\Delta$$\theta$ with $h\nu$ being the photon energy, W the work function, $\theta$ the angle with respect to the sample normal direction, the low photon energy of our VUV laser will give a better momentum resolution around the Fermi level even for the same angular resolution. For example, with W=4.3 eV, the momentum resolution $\Delta$k for $h\nu$=6.994 eV is only 0.4 times that for $h\nu$=21.2 eV.

In high resolution angle-resolved photoemission measurement, one concern is about the uniformity of the energy across the entire detector angle range that is particularly important for small energy scale measurements.  This has been examined by measuring the gold Fermi edge at low temperature. Ideally, if the angular mode is working in a perfect condition, one would expect that the Fermi edge position obtained at different detector angle should be the same. Practically this may not happen because of potential mechanical or electrical defect associated with the analyzer. Fig. 7 shows the measured two-dimensional image of the Gold Fermi edge using 30 degree angular mode (Fig. 7a) and the fitted Fermi edge position as a function of the detector angle (Fig. 7b). The deviation of the Fermi edge from its average position is less than $\pm$ 0.3 meV over the entire angular window, proving that the angular mode of our analyzer works in an excellent condition.

\subsection{Bulk sensitivity test}
The VUV laser-based ARPES is expected to greatly enhance the bulk sensitivity in probing the the electronic structure of solids. According to the ``universal curve'' of the electron escape depth versus electron kinetic energy, the electron escape depth at the present photon energy 6.994 eV is near 30 Angstrom, with a big uncertainty that can extend to 100 $\AA$\cite{Seah}. We note that this universal curve is mostly for simple metal systems, whether it is applicable to complex oxide compounds is not clear yet. While the absolute value of the electron escape depth may vary, however, one should expect enhancement of bulk sensitivity at lower photon energy even for transition metal oxides\cite{NormanEscape}.

Since it is difficult to carry out quantitative measurement of the electron escape depth, we carried out a qualitative test by measuring Bi2212 crystal sample treated at different environments. First we cleaved a nearly optimally doped Bi$_2$Sr$_2$CaCu$_2$O$_8$ (Bi2212) sample (T$_c$=90K)  {\it in situ} in the ultra-high vacuum and measured along the $\Gamma$(0,0)-Y($\pi$,$\pi$)nodal direction at a temperature of 17K; the measured result is shown in Fig. 8a. The same cleaved sample was then pulled out from the measurement chamber and exposed under 1 atmosphere N$_2$ for 1 hour in the second stage of the sample transfer system. It was then measured again along the same direction at 17K, as shown in Fig. 8b. Then the sample was pulled out and stayed in the air for 1 hour and measured again, the measured image is shown in Fig. 8c. We note that all three measurements were carried out in ultra-high vacuum condition.  It is clear that when the sample surface gets dirtier from the exposure to nitrogen and then air, the signal is getting weaker and the secondary scattering  at high binding energy is apparently getting stronger. But the  basic spectral features are still maintained in spite of such a dirty environment, a situation not possible for high photon energy like 20$\sim$50 eV.   This enhancement of bulk sensitivity will not only extend the lifetime of the sample, but also makes the surface preparation less stringent. For example, it is possible to work on some materials that are hard to cleave by mild cleaning process like annealing or sputtering.

\subsection{Space charge effect test}

One concern surrounding the utilization of laser in photoemission is the space charge effect that may originate from its high photon flux and pulsed nature\cite{ZhouSCE}.
To check on the effect, we have measured on polycrystalline gold at a low temperature ($\sim$9.2K) using different laser power. The measured Fermi level position and width as a function of laser power on the sample are plotted in Fig. 9.  The Fermi level exhibits nearly 1.5 meV increase when the laser power on the sample goes up to 130 $\mu$W (corresponding to a photon flux of 1.1$\times$10$^{14}$ photons/second, we caution that the 177.3 nm laser power here is measured using a power meter that is calibrated using 532 nm laser instead of 177.3nm laser so there could be difference) while the corresponding Fermi level width shows a broadening of less than 1 meV. This clearly demonstrates that there is a space charge effect involved in laser photoemission process. But for our VUV laser, the space charge effect is much weaker than that observed in synchrotron case\cite{ZhouSCE}.  The underlying reasons can be many-fold. First, the use of ``the quasi-CW'' laser with high repetition rate on the order of 80$\sim$100 MHz is essential for minimizing possible space-charging effects that may cause extra broadening of measured photoemission spectra\cite{ZhouSCE}. Second, the space charge effect is directly related to the number of emitted electrons. Therefore, the total number of electrons for the VUV laser is much smaller than that using high-energy synchrotron because of the reduced energy window and secondary electrons.  Third, because the space charge effect is actually balanced by the mirror charge effect, the combined effect relies on the duration of pulse. As indicated in the simulation (Fig. 14 in \cite{ZhouSCE}), the space charge effect and mirror charge effect nearly cancel each other in the range of 1$\sim$10 ps pulse duration that is the range of our VUV laser, while it grows with further increase of the duration.

\subsection{Typical example of measurement on Bi2212}


In order to test the overall performance of our UVU laser ARPES system,  we have chosen Bi2212 sample, a high temperature superconductor that has been most thoroughly studied by ARPES technique because of its layered nature and easiness of cleaving to get smooth and clean surface\cite{DamascelliReview}.  Fig. 10a  shows the raw ARPES data taken along the $\Gamma$(0,0)-Y($\pi$,$\pi$) nodal direction at a temperature of 17K, with an overall energy resolution of 1 meV.  The "kink" feature near 70 meV appears sharply in the raw data. The superior quality of the data can be best illustrated from the energy distribution curve (EDC) at the Fermi momentum (Fig. 10c) and the momentum distribution curve (MDC) at the Fermi level (Fig. 10d). The EDC width measured from our VUV Laser system for optimally-doped Bi2212 is 12 meV, and 9 meV for underdoped Bi2212 with Tc=75K, which is much sharper than that from synchrotron setup ($\sim$25meV)\cite{PashaPRL}. This is also sharper than that from the 6 eV laser ARPES measurement\cite{DessauRSI} which gives a width of $\sim$ 14 meV.  This sharp feature can be attributed to the dramatic energy resolution improvement in our VUV laser ARPES and a lower sample temperature. To our knowledge, this is the sharpest EDC along the nodal direction that has ever been reported in Bi2212 so far.  The MDC width is $\sim$0.0071 $\AA$$^{-1}$ which is also much sharper than its synchrotron counterpart(Fig. 10d)\cite{DessauRSI}.

We note that the low energy photon also brings changes in the traditional way of the data analysis. This becomes obvious in converting from the sample emission angle ($\theta$, with respect to the sample normal) to in-plane momentum k$_{||}$ (in unit of $\AA$$^{-1}$ ) which takes a form: k$_{||}$=0.5118$\sqrt{h\nu-W-E_i}$ sin$\theta$ with W being the work function of the sample and E$_i$ the binding energy. For high photon energy, like between 20$\sim$50 eV usually used in synchrotron, small uncertainty in W and neglecting E$_i$ gives rise to little change in the momentum. But for laser ARPES with lower photon energy like h$\nu$=6.994 eV, both the accuracy of W and effect of E$_i$ become pronounced. Fig. 9b shows the converted pattern considering the binding energy correction. The two momentum boundaries are no longer vertical and their spacing is shrinking with increasing binding energy.  This will have effect on EDC lineshape, and particularly on dispersion extraction. The effect will get more pronounced when the photon energy is even lower, like 6 eV UV laser\cite{DessauRSI} and should be taken into account in the data analysis in low-energy ARPES data\cite{DessauRSI,Hisor}.

\section{Discussion and Summary}

The VUV Laser has provided a novel light source for photoemission technique. Its utilization has successfully broken the barrier of 1 meV super-high energy resolution that has been a long-sought dream in photoemission technique. Compared with synchrotron radiation light source, it also provides an efficient, convenient and cheap way in a lab-based ARPES system with many unique advantages. The super-high energy resolution, high momentum resolution, super-high photon flux and enhancement of bulk sensitivity have definitely elevated the photoemission technique into a new level.

One concern related to the laser photoemission is the applicability of sudden approximation\cite{SuddenAP}. While the validation needs more theoretical and experimental efforts, we note that the measured band dispersion, Fermi surface, and overall qualitative spectral feature in Bi2212 are consistent with that measured at high photon energy. Particularly, we have observed that the EDC in Bi2212 measured at 6.994eV is even sharper than that measured at 6 eV photons\cite{DessauRSI}(Fig. 10c). This indicates that the sharpness of EDC in low energy ARPES is more likely due to dramatic improvement in the instrumental resolution, rather than the invalidity of sudden approximation.

While the VUV laser ARPES possesses a host of advantages over present synchrotron radiation source in terms of energy resolution, momentum resolution, photon flux and bulk-sensitivity, it has its limitations and further developments are needed.  The first limitation comes from its single fixed photon energy at 6.994 eV. As photoemission process involves matrix element effect where the observation of bands relies on the photon energy and polarization of the incident light\cite{BansilMatrix}, one fixed photon energy may miss some energy bands, in the worst case, may not be able to work on some materials. Development from single photon energy to tunable photon energy is necessary.  The second limitation is related to its polarization. For the same reason of matrix element effect, it is preferable to change the polarization vector of the linearly polarized light or change from linearly-polarized light into circularly polarized light. The acquisition of circularly polarized light will also help in studying magnetic materials or magnetism-related physics issues. The third limitation lies in its relatively low photon energy that leads to a small coverage of momentum space. The laser ARPES may not be able to work on some materials with small lattice constants and larger Brillouin zone.  To maintain all the advantages of the VUV laser, particularly bulk sensitivity, while reaching larger momentum space, and considering practicality of the VUV laser technology, a modest energy increase from the present 6.994eV to 7.5$\sim$8 eV is preferable and feasible.

The application of VUV laser in angle-integrated photoemission\cite{ShinP} and present angle-resolved photoemission is just a start of a whole slew of potential developments in photoemission techniques. It is expected that the super-high photon flux and super-narrow bandwidth of the VUV laser can enhance the performance of the spin-resolved photoemission, time-resolved photoemission and spatially-resolved photoemission techniques in a similarly profound way. For example, the present spin-resolved photoemission system suffers from poor energy resolution ($\sim$100meV) and most of them are operated in angle-integrated mode\cite{SpinReference}. This is mainly because the spin-detector used has a very low efficiency and the light sources used do not have enough photon flux. To get decent spin-resolved signal, one has to increase the photon flux at the expense of energy resolution, and using angle-integrated mode to further increase the signal. Compared with the synchrotron radiation where the energy resolution and the photon flux are contradictory, the super-high photon flux and super-high energy resolution of the VUV laser will make it an ideal source for spin-resolved photoemission, even with an angle-resolved capability.

In summary, we have developed the first VUV laser-based angle-resolved photoemission system with a super-high energy resolution better than 1 meV. In addition, it possesses other unique advantages like high momentum resolution, super-high photon flux and much enhanced bulk sensitivity. This has elevated the ARPES technique into a new higher level, providing a powerful tool in materials science and physics. Further development of the VUV laser technology is in progress and this novel VUV laser source will bring significant development in many other photoemission-related techniques.
\\

This work is supported from the Scientific Apparatus Development and Upgrade Project of Chinese Academy of Sciences (CAS) (Grants No. Y200420), and the Knowledge Innovation program of CAS(Grant No. KJCX2.XW.W09, KJCX2.Y200420, KJCX2.YW.H03).  XJZ also acknowledges financial support from The Hundred Talents Project of CAS and Outstanding Youth Funding from NSFC (Grant No. 10525417). We thank Prof. Zhongxian Zhao for his constant guidance and help and Mr. Weijing Yang for his excellent management in this project.  We thank John Pepper, Wangli Yang and Zahid Hussain at the ALS, Lawrence Berkeley National Lab, and Donghui Lu and Zhi-xun Shen in Stanford University for their assistance in developing this system, and B. Wannberg, Mingqi Wang in their effort in constantly working and improving the angular performance of our Scienta R4000 analyzer.

*). Corresponding author, XJZhou@aphy.iphy.ac.cn

\newpage

\begin{figure}[tbp]
\begin{center}
\includegraphics[width=0.80\columnwidth,angle=0]{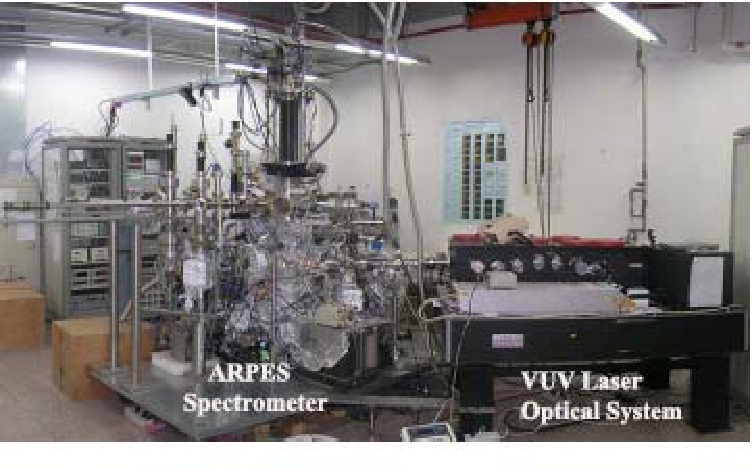}
\end{center}
\caption{ A picture of the VUV laser-based  ARPES system which is composed of two main parts: VUV laser optical system and angle-resolved photoemission spectrometer.
}
\label{Figure1}
\end{figure}

\begin{figure}[tbp]
\begin{center}
\includegraphics[width=0.8\columnwidth,angle=0]{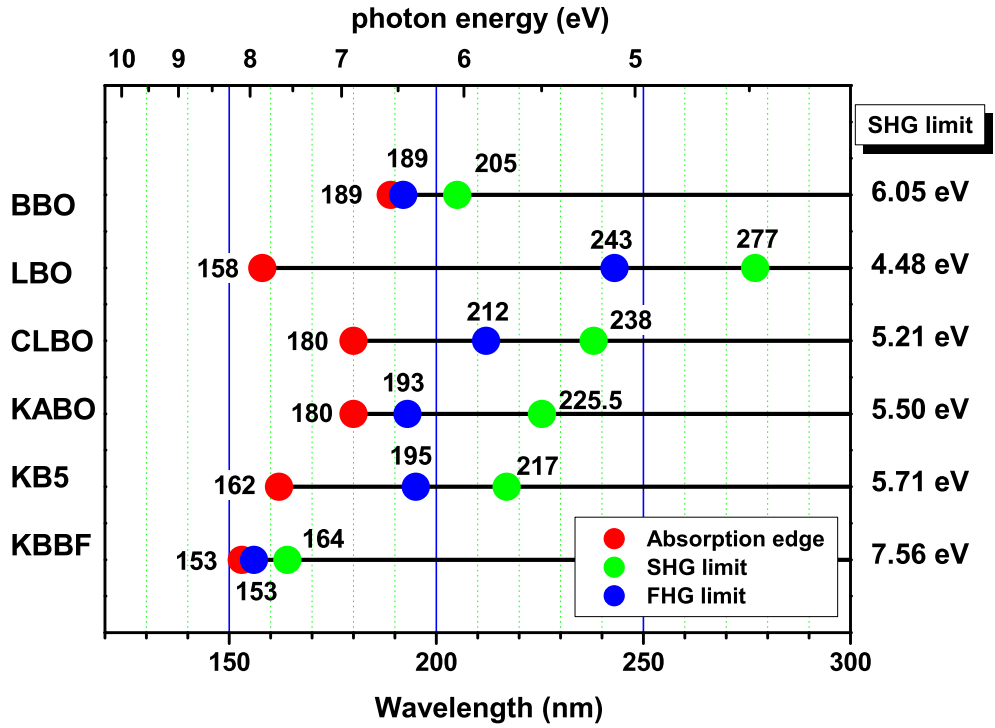}
\end{center}
\caption{A collection of some typical NLO crystals and their SHG limit, FHG (Fifth Harmonic Generation) limt and absorption edge.  The KBBF crystal shows the shortest SHG wavelength (highest SHG energy) among all the available NLO crystals.
}
\label{Figure2}
\end{figure}

\begin{figure}[tbp]
\begin{center}
\includegraphics[width=0.8\columnwidth,angle=0]{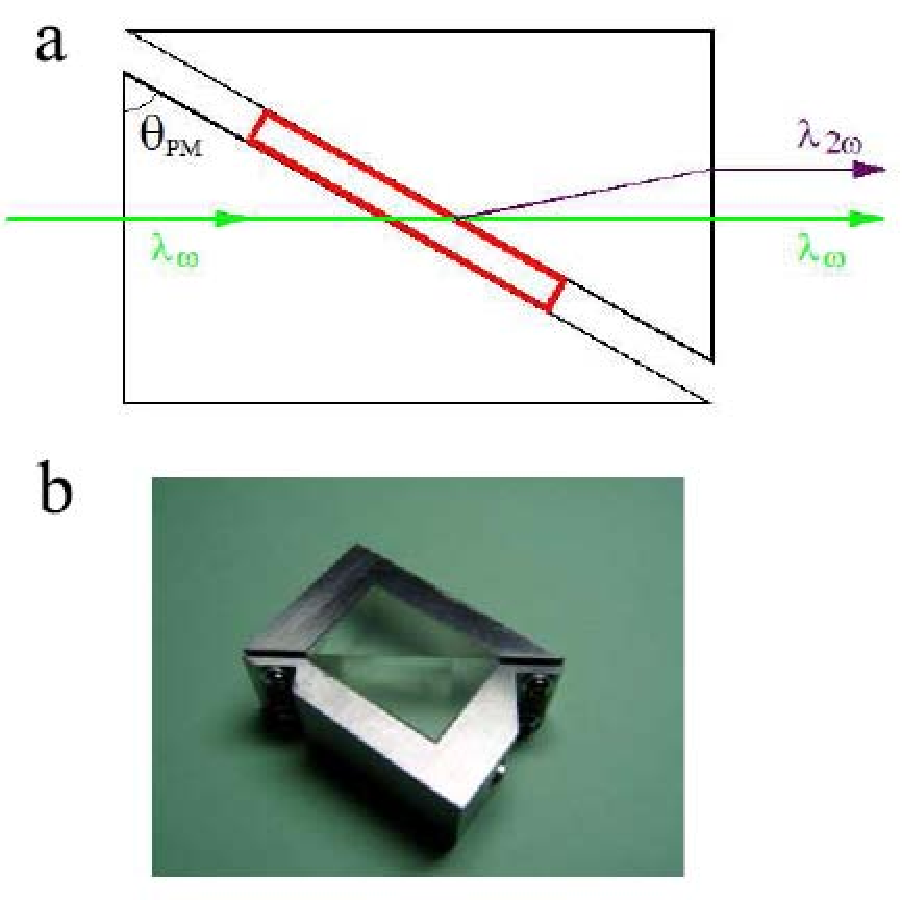}
\end{center}
\caption{(a).The schematic sandwich structure of an optically contacted,
prism-coupled KBBF crystal device; (b). A photo of a typical KBBF-CaF$_2$ prism-coupled device.
}
\label{Figure3}
\end{figure}

\begin{figure}[tbp]
\begin{center}
\includegraphics[width=0.80\columnwidth,angle=0]{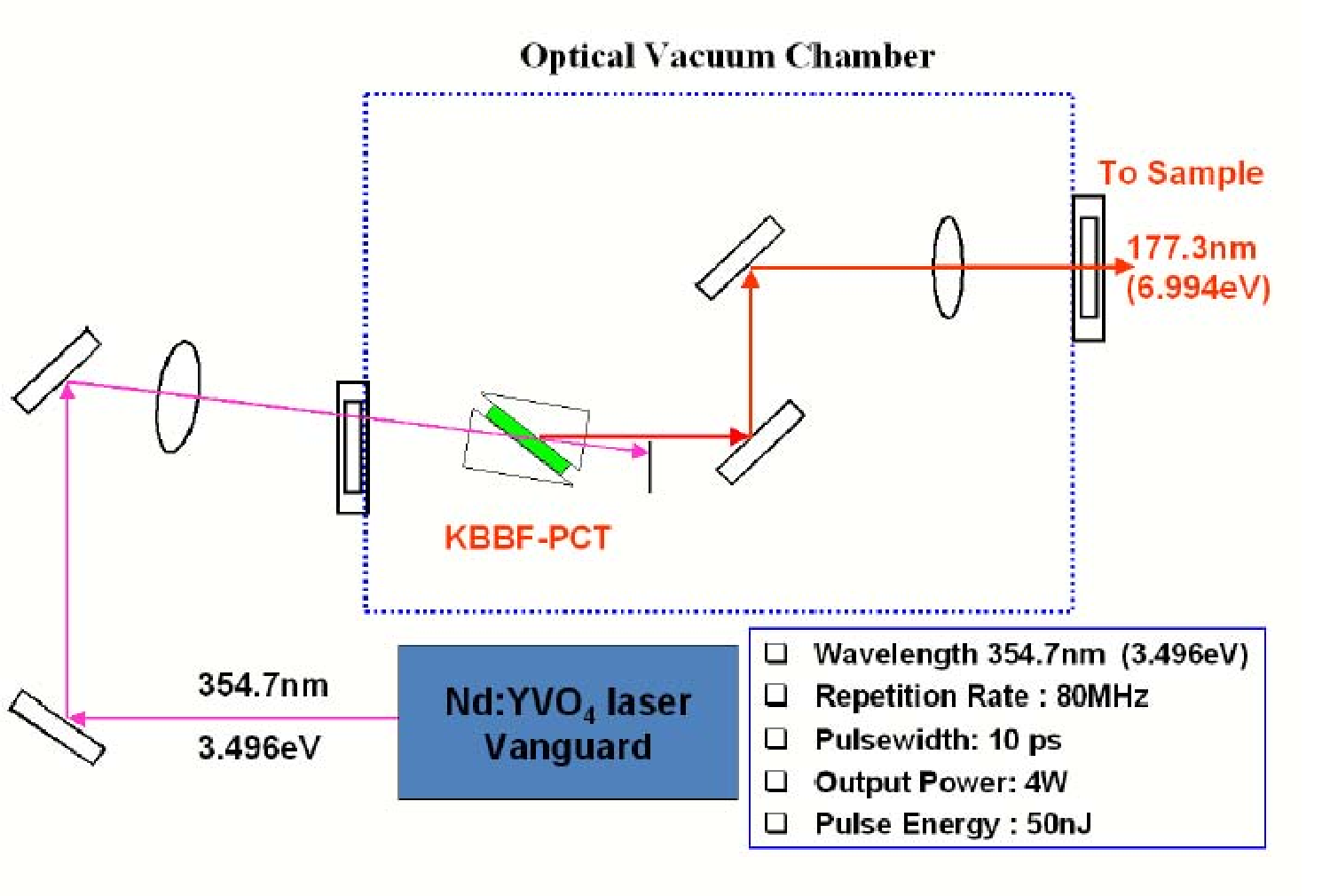}
\end{center}
\caption{A schematic layout of our VUV laser optical system.
}
\label{Figure4}
\end{figure}

\begin{figure}[tbp]
\begin{center}
\includegraphics[width=0.80\columnwidth,angle=0]{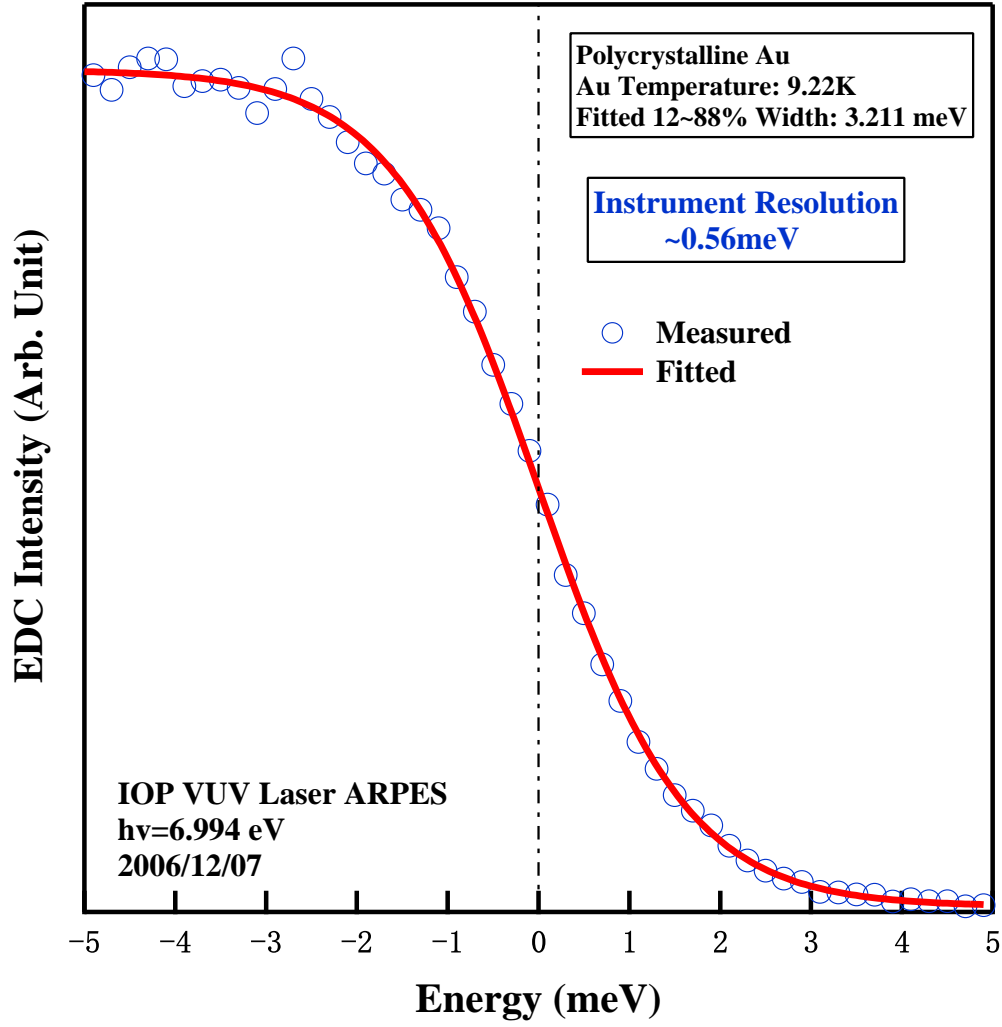}
\end{center}
\caption{Energy resolution test of our VUV laser ARPES system by measuring on a clean polycrystalline gold at a temperature of 9.22 K.  The measured data (blue open circles) are fitted by the Fermi function (red solid line) and the overall 12$\sim$88$\%$ linewidth obtained is 3.211 meV. After subtracting the contribution from the thermal broadening at 9.22K, the instrumental energy resolution is 0.56 meV.
}
\label{Figure5}
\end{figure}

\begin{figure}[tbp]
\begin{center}
\includegraphics[width=0.70\columnwidth,angle=0]{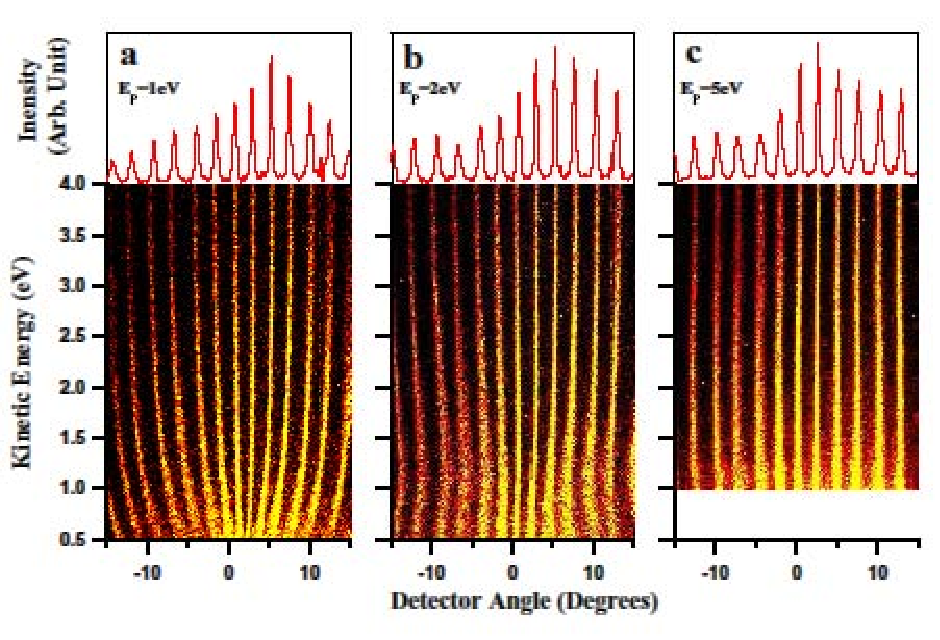}
\end{center}
\caption{The angular testing results of our R4000 electron energy analyzer using the Scienta wire-and-slit device. The measured angular distribution patterns at
(a). 1 eV, (b). 2 eV and (c). 5 eV pass energy are shown at the bottom panels. The corresponding line profiles near 2.7 eV kinetic energy are shown on the top panels.
}
\label{Figure6}
\end{figure}

\begin{figure}[tbp]
\begin{center}
\includegraphics[width=0.80\columnwidth,angle=0]{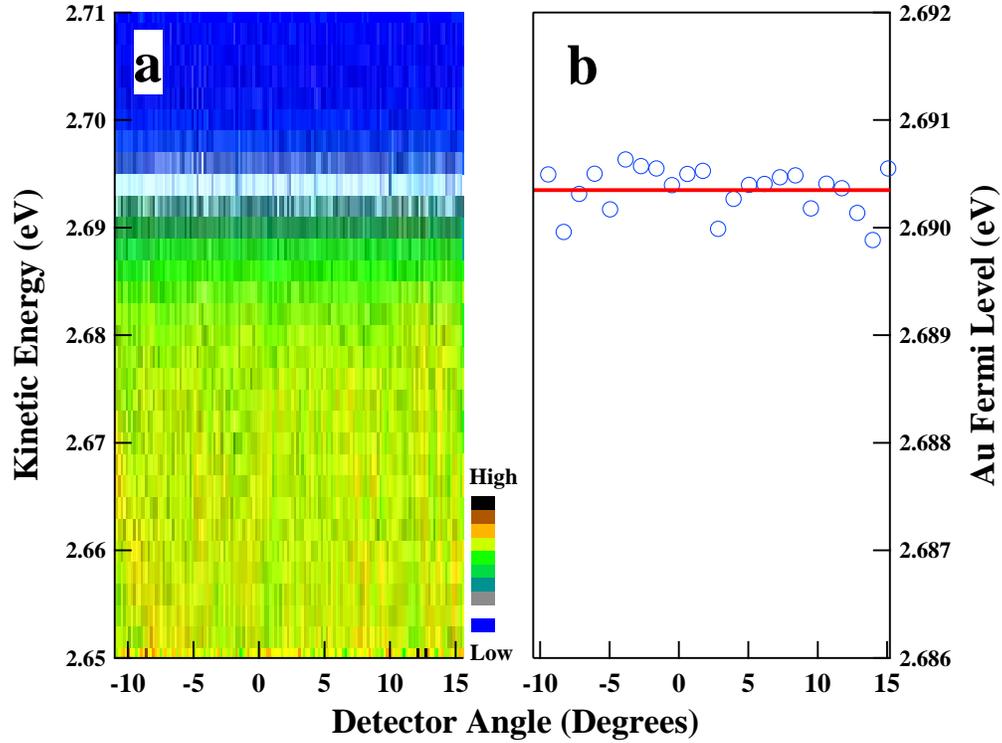}
\end{center}
\caption{Fermi level uniformity test for the angular mode of our electron energy analyzer.  (a). Two-dimensional photoemission image taken on a clean polycrystalline gold sample at a low temperature using the angular mode of 30 degrees and the pass energy of 2 eV. (b). The corresponding Fermi level position as a function of the detector angle. Over the entire detection window, The variation of the Fermi level position falls within $\pm$0.3meV of the average value.
}
\label{Figure7}
\end{figure}

\begin{figure}[tbp]
\begin{center}
\includegraphics[width=0.80\columnwidth,angle=0]{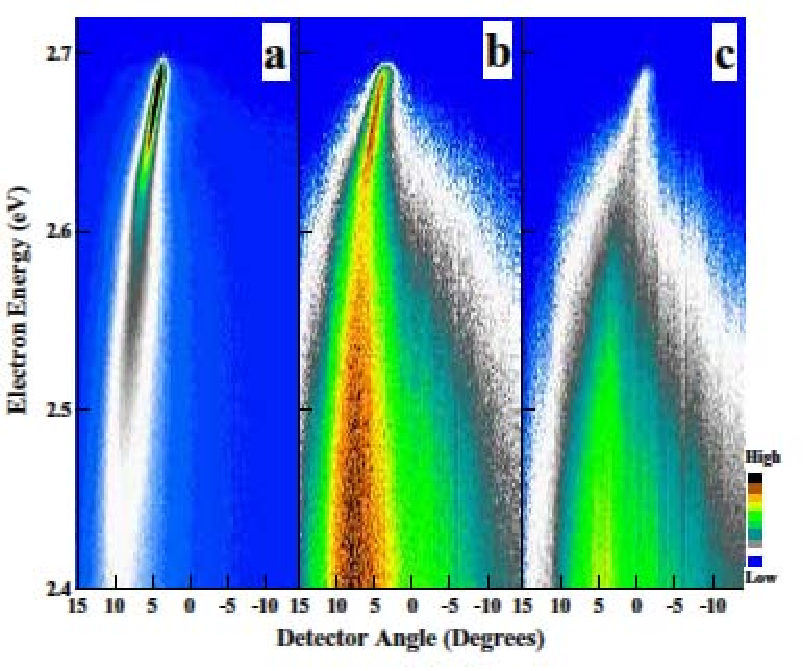}
\end{center}
\caption{Bulk sensitivity test on a typical Bi2212 sample. All data are taken under ultra-high vacuum better than 1$\times$10$^{-10}$ mbar at a tempearture of 15 K along the (0,0)-($\pi$,$\pi$) nodal direction. (a). Photoemission data measured on the Bi2212 surface {\it in situ} cleaved in UHV chamber; (b).The same cleaved surface after being exposed in 1 atm N$_2$ for 1 hour;  (c).The same surface after being further exposed in 1 atm air for 1 hour.
}
\label{Figure8}
\end{figure}

\begin{figure}[tbp]
\begin{center}
\includegraphics[width=0.80\columnwidth,angle=0]{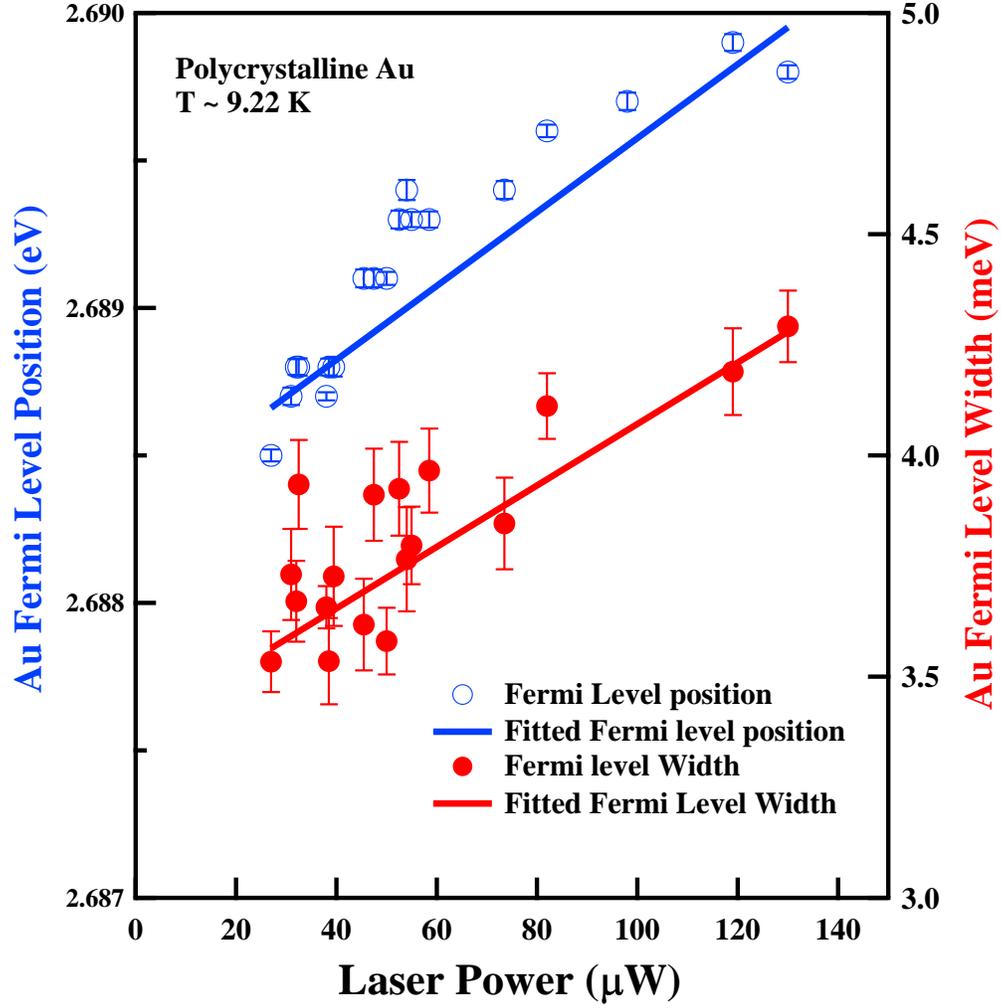}
\end{center}
\caption{Space charge effect test of the VUV laser ARPES system by measuring Fermi level position and width on a polycrystalline gold at a temperature of 9.22K using different laser power. The measured Fermi edges are fitted by Fermi function and the obtained Fermi level position (blue empty circles) and width (12$\sim$88$\%$) (red solid circles) are fitted by linear lines as solid blue line and solid red line, respectively.
}
\label{Figure9}
\end{figure}

\begin{figure}[tbp]
\begin{center}
\includegraphics[width=0.90\columnwidth,angle=0]{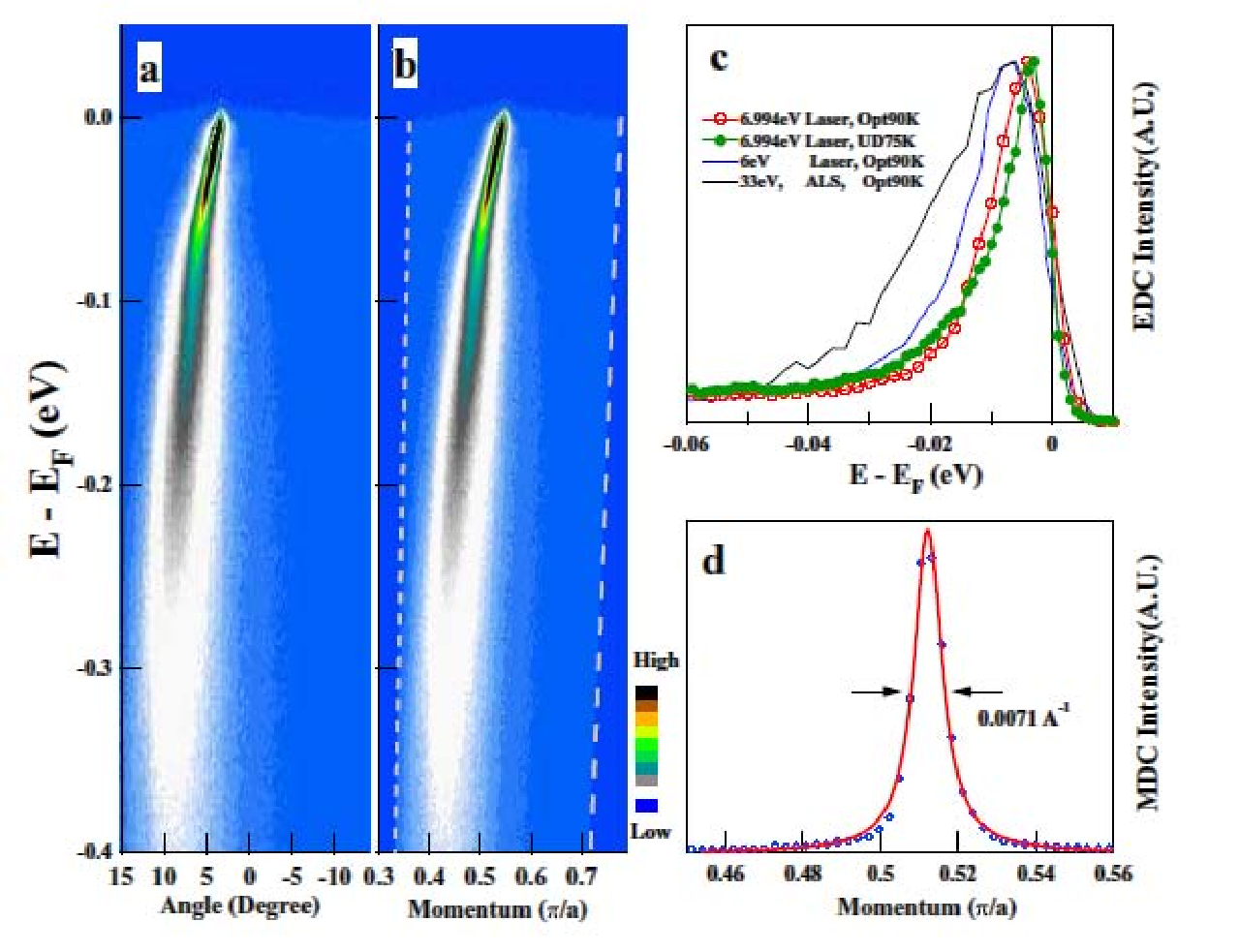}
\end{center}
\caption{High resolution ARPES data taken on Bi2212 using our VUV laser ARPES system. The data were taken along the (0,0)-($\pi$,$\pi$) nodal direction at a temperature of 15 K and an overall instrumental resolution of 1 meV. (a) Raw data showing the photoelectron intensity (represented by false color) as a function of electron energy and detector angle.  (b). The same image as (a) but the angle is converted into momentum. (c). The EDC spectra at the Fermi momentum measured using our VUV laser ARPES system on optimally doped Bi2212 (T$_c$=90K, red circle and line), and underdoped Bi2212 (T$_c$=75 K, green circle and line).  For comparison, the EDCs measured from the 6 eV UV laser ARPES (blue line)\cite{DessauRSI} and from synchrotron system at 33 eV (black line) on the optimally doped Bi2212\cite{PashaPRL} are also plotted.   (d) The MDC spectrum taken at the Fermi energy from (b) on optimally doped Bi2212.
}
\label{Figure9}
\end{figure}


\begin{thebibliography}{99}

\bibitem{Huefner} S. Huefner, Photoelectron Spectroscopy: Principles and Applications (Springer-Verlag, Berlin, 1995).

\bibitem{SKevan} Angle-Resolved Photoemission: Theory and Current Applications, edited by S. D. Kevan, (Elsevier, The Netherlans, 1992).

\bibitem{EinsteinPE} A. Einstein, Ann. Physik {\bf 17}, 132 (1905).

\bibitem{SFunction} L. Hedin and S. Lundqvist, in Solid State Physics, edited by F. Seitz, D. Turnbull and H. Ehrenreich, Aca-demic Press (1969).

\bibitem{DamascelliReview} A. Damascelli, Z. Hussain, and Z.-X. Shen, Rev. Mod. Phys. {\bf 75}, 473 (2003).

\bibitem{CampuzanoReview} J. C. Campuzano, M. R. Norma and M. Randeria, in Physics of Superconductors, Vol. II, ed. K. H. Benne-mann and J. B. Ketterson (Springer, Berlin, 2004), pp.167-273.

\bibitem{ZhouReview} X.J. Zhou et al., chapter 3, Handbook of High Temperature Superconductivity, Edited by J. R. Schrieffer, Springer-Verlag (2006).

\bibitem{SpecialJESRP} Special issue of J. Electron Spectroscopy and Related Phenomena, {\bf {117-118}} 1 (2001).

\bibitem{NSmith} N. V. Smith, CRC Crit. Rev. Solid State Sci. {\bf 2}, 45 (1971).

\bibitem{BednorzMuller} J. G. Bednorz and K. A. Muller, Zeitschrift fur Physik B {\bf 64},189 (1986).

\bibitem{ScienceSpecial} Special issue of Science {\bf288}, No. 5465, (2000).

\bibitem{ScientaWebSpecsWeb} See, for example,  the homepage of VG Scienta AB (http://www.vacgen.com) and the homepage of SPECS GmbH (http://www.specs.de).

\bibitem{MDCMethod} P. Aebi et al., Phys. Rev. Lett. {\bf 72}, 2757 (1994); T. Valla, A. V. Fedorov, P. D. Johnson, B. O. Wells, S. L. Hulbert, Q. Li, G. D. Gu, and N. Koshizuka, Science {\bf285}, 2110 (1999).

\bibitem{HuefnerPbShinPb} F. Reinert, B. Eltner, G. Nicolay, D. Ehm, S. Schmidt, and S. Huefner, Phys. Rev. Lett. {\bf 91}, 186406 (2003); A. Chainani, T. Yokoya, T. Kiss, and S. Shin, Phys. Rev. Lett. {\bf 85},1966(2001).

\bibitem{ZhouPRL} X. J. Zhou et al., Phys. Rev. Lett. {\bf 95}, 117001(2005).

\bibitem{Seah}M. P. Seah and W. A. Dench, Surf. Interface Anal. {\bf 1}, 2 (1979).

\bibitem{Suga} A. Sekiyama, T. Iwasaki, K. Matsuda, Y. Saitoh, Y. Onuki, and S. Suga, Nature {\bf 403}, 396 (2000).

\bibitem{ZhouSCE}X. J. Zhou et al., J. Electron Spectroscopy and Related Phenomena {\bf 142}, 27 (2005).

\bibitem{SuddenAP} L. Hedin and J. D. Lee, J. Electron Spectroscopy and Related Pheonomena {\bf 124}, 289 (2002).

\bibitem{ShinP} T. Kiss et al., Phys. Rev. Lett. {\bf 94}, 057001 (2005).

\bibitem{DessauRSI}J. D. Koralek et al., Rev. Sci. Instrum. {\bf 78}, 053905 (2007); J. D. Koralek et al., Phys. Rev. Lett. {\bf 96}, 017005 (2006).

\bibitem{NoteVUV} Here the VUV (UV) is defined by the  photon energy larger (smaller)
than 6.5eV. VUV laser will be absorbed in air, so it can only pass through vacuum or some inert gases.

\bibitem{BBORef}C T. Chen et al., Sci. Sin. B {\bf 28}, 235 (1985).

\bibitem{LBORef}C. T. Chen et al., J. Opt. Sot. Am. {\bf B 6}, 616 (1989).

\bibitem{ChenRef} C. T. Chen et al., Appl. Phys. Lett. {\bf 68}, 2930 (1996).

\bibitem{Chen2002} C. T. Chen et al., Optics Lett. {\bf 27}, 637 (2002).

\bibitem{Togashi} T. Togashi et al., Optics Lett. {\bf 28}, 254 (2003).

\bibitem{ParameterRef} C. Chen, Z. Lin and Z. Wang, Appl. Phys. {\bf B 80}, 1 (2005).

\bibitem{Hisor} This is realized recently in synchrotron radiation facilities with low sorage ring energy that are good at producing low energy photons, like, e.g., Hiroshima Synchrotron Radiation Center at http://www.hsrc.hiroshima-u.ac.jp/. An example of related work: T. Yamasaki et al., Phys. Rev. {\bf B 75}, 140513 (2007).

\bibitem{NormanEscape}M. Norman et al., Phys. Rev. {\bf B59}, 11191 (1999).

\bibitem{PashaPRL} P. V. Bogdanov et al., Phy. Rev. Lett. {\bf B 85}, 2581 (2000).

\bibitem{BansilMatrix} A. Bansil and M. Lindroos, Phys. Rev. Lett. {\bf 83}, 5154 (1999).

\bibitem{SpinReference} J.-H. Park et al., Nature {\bf 392}, 794 (1998);
S. W. Yu et al., Phy. Rev. {\bf B 73}, 75116 (2006);
W. H.Wang et al., Phy. Rev. {\bf B 71} 144416 (2005);
C. De Nadii et al., Phy. Rev. {\bf B 68} 212401 (2003);
M. Sawada et al., Phy. Rev. {\bf B 63} 195407 (2001).
D. J. Huang et al., Rev. Sci. Instru. {\bf 73} 3778 (2002).


\end{thebibliography}
\end{document}